\documentclass[final,aps,prb,preprint,superscriptaddress,showpacs]{revtex4-1}

\usepackage{graphicx}
\usepackage{dcolumn}
\usepackage{amsmath,bm}
\usepackage{color}%
\usepackage{multirow}
\usepackage{array} 
\usepackage{esvect}
\usepackage{braket}
\usepackage{makecell}
\usepackage[utf8]{inputenc}
\usepackage{kotex}
\usepackage{xr}
\usepackage{soul,xcolor}
\setstcolor{red} 

\usepackage{xr}

\makeatletter
\newcommand*{\addFileDependency}[1]{
  \typeout{(#1)}
  \@addtofilelist{#1}
  \IfFileExists{#1}{}{\typeout{No file #1.}}
}
\makeatother

\newcommand*{\myexternaldocument}[1]{%
    \externaldocument{#1}%
    \addFileDependency{#1.tex}%
    \addFileDependency{#1.aux}%
}

\myexternaldocument{SI}  

\begin{document}

\title{GaN Nucleation Landscape on Patterned Sapphire Shaped by the Growth Temperature of Directly Grown Boron-Compound Masks}

\author{Yunjin Heo}
\affiliation{Department of Physics and Research Institute for Basic Sciences, Kyung Hee University, Seoul, 02447, Republic of Korea}

\author{Donghoi Kim}
\affiliation{Department of Information Display, Kyung Hee University, Seoul, 02447, Republic of Korea}

\author{Hyeonoh Jo}
\affiliation{Department of Physics and Research Institute for Basic Sciences, Kyung Hee University, Seoul, 02447, Republic of Korea}

\author{Aelim Ha}
\author{Soohyung Park}
\affiliation{Advanced Analysis Center, Korea Institute of Science and Technology (KIST), Seoul 02792, Republic of Korea}

\author{Jaewu Choi}
\affiliation{Department of Information Display, Kyung Hee University, Seoul, 02447, Republic of Korea}

\author{Chinkyo Kim}
\email{ckim@khu.ac.kr}
\affiliation{Department of Physics and Research Institute for Basic Sciences, Kyung Hee University, Seoul, 02447, Republic of Korea}
\affiliation{Department of Information Display, Kyung Hee University, Seoul, 02447, Republic of Korea}


\begin{abstract}
The growth temperature of directly grown boron-compound masks on patterned sapphire can modify the local accessibility of the underlying sapphire surface and thereby alter the subsequent nucleation behavior of GaN. In this work, we investigate how ammonia-borane-derived boron-compound masks grown at different temperatures shape the GaN nucleation landscape within circular SiO$_2$ openings during the initial stage of epitaxial lateral overgrowth. The preferential nucleation position of GaN changes systematically with mask growth temperature: masks grown at 700--750$^\circ$C produce pronounced edge-biased distributions, whereas higher-temperature masks lead to more inward-shifted and spatially sparse GaN domains. Quantitative analysis of the GaN areal fraction, the number of visibly isolated domains, and the radial distribution of domain centers shows that the mask growth temperature affects both the amount of GaN coverage and the spatial arrangement of GaN domains within each opening. The nonmonotonic change in the number of visibly isolated domains is interpreted as a consequence of competition between reduced lateral merging and reduced effective substrate accessibility, rather than as a direct measure of the number of active nucleation sites. Kinetic Monte Carlo simulations reproduce the essential experimental trends by varying the effective density and radial distribution of substrate-accessible sites. These results suggest that the growth temperature of directly formed boron-compound masks provides a practical means of reshaping the intra-opening GaN nucleation landscape by controlling the spatial distribution and effectiveness of local pathways through which GaN precursors can access the underlying sapphire surface.
\end{abstract}



\maketitle

\section{Introduction}

Hexagonal boron nitride (hBN) and related boron nitride materials have attracted considerable attention as two-dimensional insulating materials for electronic, optoelectronic, and epitaxial applications.\cite{Ma-Nature-606-88,Abidi-AOM-7-1900397,Castelletto-BJN-11-740} Owing to their wide band gap, high thermal and chemical stability, and atomically smooth surfaces with few dangling bonds, BN-based films can provide chemically stable and weakly interacting interfaces for the growth of other semiconductor materials.\cite{Molaei-ACS-3-5165,Maestre-JPM-4-044018,Ma-AMR-3-748} In particular, the van der Waals nature of hBN interfaces can relax lattice-matching constraints and has therefore been explored for the epitaxial growth and release of III-nitride semiconductors, including GaN.\cite{Ravi-ACSAEM-5-146,Xu-CGD-23-2196-r,Wu-SR-6-34766} However, because atomically smooth and dangling-bond-free hBN surfaces provide few native nucleation sites, uniform GaN nucleation on hBN remains challenging and often requires defects, surface treatments, wetting or nucleation layers, or locally accessible regions through which precursor adsorption and nucleation can be initiated.\cite{Ravi-ACSAEM-5-146,Xu-CGD-23-2196-r}

BN films have been synthesized by various methods, including molecular beam epitaxy (MBE),\cite{Liu-APL-116} sputtering,\cite{Abbas-ML-227-284,Chen-ACS-14-7004} and chemical vapor deposition (CVD).\cite{Chen-JMC-28-14341,Lee-SR-9-10590} Among these, CVD is particularly attractive for the scalable growth of large-area BN-based films.\cite{Tay-JMCC-2-1650} Previous studies have shown that growth conditions such as precursor chemistry,\cite{Yang-JCG-482-1,Chen-KEM-843-90,Bansal-JMR-36-4678} carrier gas,\cite{Kim-AIP-7} growth temperature,\cite{Ahmed-MRE-4-015007,Liu-CGD-24-810} and reactor configuration\cite{Sutorius-16-15782,Saha-AIP-11} strongly influence the resulting film properties, including growth rate,\cite{Zhu-CGD-23-7662,Rice-JCG-485-90} electrical characteristics,\cite{Sattari-ACS-15-7274} structural quality,\cite{Stehle-CM-24-8041} band gap,\cite{Durandurdu-JNCS-427-41} and dielectric constant.\cite{Hong-Natrue-582-511-r} These results indicate that the growth conditions of BN-based films are not merely synthesis parameters, but can determine their subsequent functionality as interlayers or epitaxial masks.

For epitaxial-mask applications, the most relevant property is not necessarily the formation of phase-pure, highly crystalline hBN alone. Instead, the key issue is how the film modifies the local accessibility of the underlying substrate during subsequent epitaxial growth. When a BN-based or boron-compound film is directly formed on a patterned substrate, its local coverage, thickness, structural continuity, and defect distribution can create spatially nonuniform substrate-accessible regions. These regions may correspond to nanoscale openings, percolative through-pathways, locally thinner parts of the film, or discontinuous mask coverage. Such local variations are especially important for SiO$_2$-patterned sapphire substrates, where GaN nucleation occurs inside lithographically defined circular openings and the position of the initial nuclei can strongly influence the subsequent growth morphology.

Despite extensive studies on high-temperature BN growth and its use as a van der Waals epitaxy template, relatively low-temperature, precursor-derived boron-compound films and their function as epitaxial masks have been much less explored. This distinction is important because low-temperature growth may produce films with incomplete crystallinity, mixed bonding states, or spatially nonuniform coverage, rather than ideal hBN. At the same time, such imperfect or partially continuous films may be useful as functional masks because they can provide local pathways through which GaN precursors access the underlying sapphire surface. Therefore, understanding how the growth temperature of directly formed boron-compound masks affects the effective accessibility of sapphire is essential for controlling GaN nucleation on patterned substrates.

In patterned substrates, this question becomes a spatial problem. If the substrate-accessible regions in the boron-compound mask are distributed nonuniformly within each circular opening, GaN nucleation should not occur uniformly over the opening area. Instead, the preferred nucleation position, GaN areal fraction, and number of visible domains may change depending on the spatial distribution and effectiveness of these accessible sites. Importantly, such behavior does not need to originate from a single microscopic mechanism. It may arise from a redistribution of substrate-accessible pathways, from radially nonuniform film coverage, or from local thinning or discontinuity of the boron-compound mask near the SiO$_2$ boundary. The growth temperature of the mask may therefore provide a route to reshaping the GaN nucleation landscape by modifying the effective accessibility of the sapphire surface, rather than simply changing the overall nucleation probability.

In this study, we investigate how ammonia-borane-derived boron-compound masks grown at different temperatures influence GaN nucleation on SiO$_2$-patterned sapphire substrates. Because the films are formed at relatively low temperatures and their phase purity cannot be assumed a priori, we conservatively refer to them as boron-compound masks unless specific chemical or structural evidence is discussed. By combining control experiments, SEM-based domain statistics, radial-position analysis, X-ray diffraction, X-ray photoelectron spectroscopy, and kinetic Monte Carlo simulations, we show that the mask growth temperature systematically changes the GaN nucleation landscape within circular openings. The results suggest that this behavior is associated with temperature-dependent changes in the spatial distribution of effective substrate-accessible sites, including possible through-pathways, local thinning, and discontinuities in the boron-compound mask. This provides a practical approach for controlling GaN nucleation on patterned sapphire through the direct growth of boron-compound masks.  

Previous studies on GaN growth on hBN have shown that defects, wrinkles, plasma-induced dangling bonds, or atomic steps in transferred hBN layers can act as preferential nucleation sites for GaN and promote the formation of continuous or exfoliable GaN films.\cite{Ravi-ACSAEM-5-146,Xu-CGD-23-2196-r} In contrast, the present work focuses on directly grown ammonia-borane-derived boron-compound masks on lithographically patterned sapphire, without relying on transferred hBN wrinkles or plasma-induced activation of the hBN surface. We show that the mask growth temperature itself reshapes the intra-opening GaN nucleation landscape by modifying the spatial distribution of effective substrate-accessible regions within circular SiO$_2$ openings.

Recent studies have demonstrated several routes for controlling GaN nucleation using two-dimensional masks, including TBA-assisted spin-coated rGO nanosieves, self-adjusting solution-processed h-BN flake stacks, and O$_2$-plasma-perforated graphene masks.\cite{Beak-CGD-25-7557,Ha-CGD-25-8431,An-CGD-25-10571} These approaches control precursor access primarily by engineering the morphology, stacking, or perforated-area fraction of preformed or post-processed 2D-material masks. The present work addresses a different process variable: the growth temperature of an ammonia-borane-derived boron-compound mask that is formed directly on SiO$_2$-patterned sapphire. Rather than introducing perforations by plasma treatment or tuning the stacking of spin-coated flakes, we show that the mask growth temperature itself reshapes the intra-opening GaN nucleation landscape by modifying the spatial distribution and effectiveness of local substrate-accessible regions.

\section{Experimental}

\subsection{Growth of boron-compound masks and GaN}

A 50-nm-thick SiO$_2$ layer was deposited on a \(c\)-plane sapphire substrate and subsequently patterned by conventional photolithography and etching to define circular openings. The openings had a diameter of 4 \(\mu\)m and a center-to-center spacing of 7 \(\mu\)m. Ammonia-borane-derived boron-compound masks were then directly grown on the SiO$_2$-patterned \(c\)-plane sapphire substrates at different temperatures for 15 min. The mask-growth process was carried out at a reactor pressure of 0.7 Torr with H$_2$ and Ar flow rates of 10 sccm and 0.1 slm, respectively. For the main GaN nucleation-position and domain-statistics analyses, the mask growth temperature was varied from 550 to 850$^\circ$C.  Additional boron-compound mask samples grown at 450$^\circ$C were included only for chemical-state comparison by XPS and were not used in the main SEM-based GaN nucleation statistics.

After mask formation, GaN was grown by hydride vapor phase epitaxy at 957$^\circ$C and atmospheric pressure using HCl and NH$_3$ flow rates of 10 sccm and 0.5 slm, respectively. The GaN growth time was fixed at 21 s for all samples in this series in order to isolate the effect of the boron-compound mask growth temperature on the GaN nucleation behavior.

Additional GaN/boron-compound/sapphire samples prepared under similar, but not identical, growth conditions (growth duration of 1 minute) were used for complementary X-ray diffraction analysis. These XRD measurements were intended to verify whether GaN grown in this material system can retain crystallographic alignment with the underlying sapphire substrate, rather than to provide a one-to-one structural characterization of the exact samples used for the SEM-based nucleation statistics.

\subsection{Analysis of radial distribution of GaN domains}

For the radial-position analysis of GaN domains, SEM images were first segmented using manually prepared binary masks, where GaN domains were assigned as the foreground and the surrounding region as the background. The center and radius of each circular SiO$_2$ opening were determined from the SEM image using Fiji/ImageJ.\cite{Schindelin-NM-9-676} To avoid uncertainty associated with the opening boundary, an inner analysis radius \(R_{\mathrm{inner}}\) was defined by subtracting a fixed boundary margin from the measured opening radius. Small isolated artifacts in the binary mask were removed before further analysis.

The positions of isolated domains or local nucleation centers within merged domains were then estimated from the cleaned binary mask. For this purpose, a Euclidean distance transform was applied to the GaN-domain mask, and local maxima in the distance map were identified as estimated domain-center or seed positions. When multiple local maxima were present within a connected GaN region, watershed segmentation based on the distance transform was used to separate the merged region into individual domain-associated regions.\cite{Legland-Bioinformatics-32-3532} For each estimated domain center, the radial distance \(r\) from the center of the circular opening was calculated and normalized by \(R_{\mathrm{inner}}\). The resulting normalized coordinate \(r/R_{\mathrm{inner}}\), where \(r/R_{\mathrm{inner}}=0\) corresponds to the opening center and \(r/R_{\mathrm{inner}}=1\) corresponds to the inner boundary of the SiO$_2$ opening, was used to construct the radial distribution histograms.

To visualize the overall trend of the radial distributions, histograms of \(r/R_{\mathrm{inner}}\) were plotted using a fixed bin number for all samples. The radial distributions were used as position-resolved descriptors of the GaN domain distribution within each circular opening.

Because this analysis is based on the final SEM morphology after finite lateral growth, the extracted positions are referred to as estimated domain-center positions rather than direct observations of the initial atomic-scale nucleation sites. Nevertheless, for isolated or weakly merged domains, these positions can serve as a practical proxy for the spatial distribution of effective substrate-accessible sites, including local thinning, discontinuities, or through-pathways in the boron-compound mask.

\subsection{Kinetic Monte Carlo simulation}

Kinetic Monte Carlo-type simulations were carried out to evaluate how the effective density and radial distribution of substrate-accessible sites in the boron-compound mask affect GaN nucleation within a circular opening. The opening was modeled as a disk-shaped region on a two-dimensional \(601 \times 601\) grid. The opening radius was set to 0.70 of the half-width of the simulation box, and the GaN-covered area fraction was evaluated only within this circular opening, although lateral growth was allowed to extend outside the opening. The simulation follows a coarse-grained stochastic nucleation-and-growth framework, in which nuclei are generated randomly at allowed sites and subsequently grow laterally until impingement, analogous to classical nucleation-and-growth descriptions.\cite{Johnson-TAIMME-135-415,Avrami-JCP-7-1103}

GaN nucleation was allowed only at effective substrate-accessible sites. The effective accessible-site fraction, \(f_{\mathrm{acc}}\), was used as a coarse-grained parameter representing the fraction of grid points within the opening that permit precursor access to the sapphire surface. This parameter was not treated as a directly measured microscopic opening fraction. The values of \(f_{\mathrm{acc}}\) were set to 0.77, 0.45, 0.165, 0.085, 0.080, and 0.025 for the representative 550, 650, 700, 750, 800, and 850$^\circ$C cases, respectively. The values of \(f_{\mathrm{acc}}\) were chosen to qualitatively reproduce the experimentally observed decrease in GaN areal fraction with increasing mask growth temperature, while keeping the intrinsic nucleation and growth parameters fixed.  For the 550, 650, 800, and 850$^\circ$C cases, the accessible sites were randomly and uniformly distributed within the circular opening. For the 700 and 750$^\circ$C cases, a ring-biased accessible-site distribution was introduced near the SiO$_2$ opening boundary as a representative test of edge-concentrated substrate accessibility under selected conditions, rather than as a direct measurement of the microscopic mask morphology. In these cases, the target ratio of ring-region to inner-region accessible sites was set to 9:1. The ring was centered at \(0.90R_{\mathrm{open}}\), where \(R_{\mathrm{open}}\) is the opening radius, and the inner region was defined as \(r < 0.70R_{\mathrm{open}}\).

Nuclei were generated stochastically from a Poisson process at the effective substrate-accessible sites. The nucleation rate per accessible site was kept identical for all representative temperature cases. In the simulation, the time step was \(\Delta t = 0.10\), the total simulation time was \(t_{\mathrm{end}} = 12.0\), and the nucleation rate parameter was \(J_{\mathrm{acc}} = 2.14 \times 10^{-4}\) per accessible site per unit simulation time. Thus, the corresponding nucleation probability per accessible site per time step was approximately \(J_{\mathrm{acc}}\Delta t = 2.14 \times 10^{-5}\). After nucleation, lateral growth was calculated using an anisotropic kinetic-Wulff arrival-time kernel, conceptually related to arrival-time descriptions of monotonically advancing fronts.\cite{Sethian-PNAS-93-1591} The growth kernel was constructed using 360 angular directions, \(v_{\min}=1.0\), \(v_{\max}=6.0\), \(\sigma=0.08\), and an overall growth scale of \(S=2.8\). The same nucleation rate per accessible site and the same lateral-growth kernel were used for all cases. Therefore, the simulated changes in GaN coverage, domain number, and preferred nucleation position arise from changes in \(f_{\mathrm{acc}}\) and the radial distribution of effective substrate-accessible sites, rather than from changes in the intrinsic nucleation rate or growth law.

\section{Results and discussion}

\subsection{Control experiment for separating precursor-derived mask effects from thermal pretreatment}

Because the boron-compound mask in the present system is formed by introducing ammonia borane during a relatively low-temperature preparation step, it is necessary to verify whether the observed change in GaN nucleation originates from the precursor-derived surface layer itself or from the thermal pretreatment of sapphire. In particular, since the preparation temperature is relatively low compared with typical temperatures used for substantial BN film growth, one may question whether little or no boron-containing layer is formed under the present conditions.\cite{Singhal-TSF-733-138812} In addition, thermal exposure of sapphire can itself modify the surface morphology and chemical state, thereby affecting subsequent GaN nucleation.\cite{Tsuda-ASS-216-585,Curiotto-SS-603-2688,Simeonov-SS-603-232} To separate these effects, a control experiment was performed using H$_2$-annealed sapphire prepared under otherwise identical thermal conditions but without ammonia borane precursor supply.

\begin{figure}
\includegraphics[width=1.0\columnwidth]{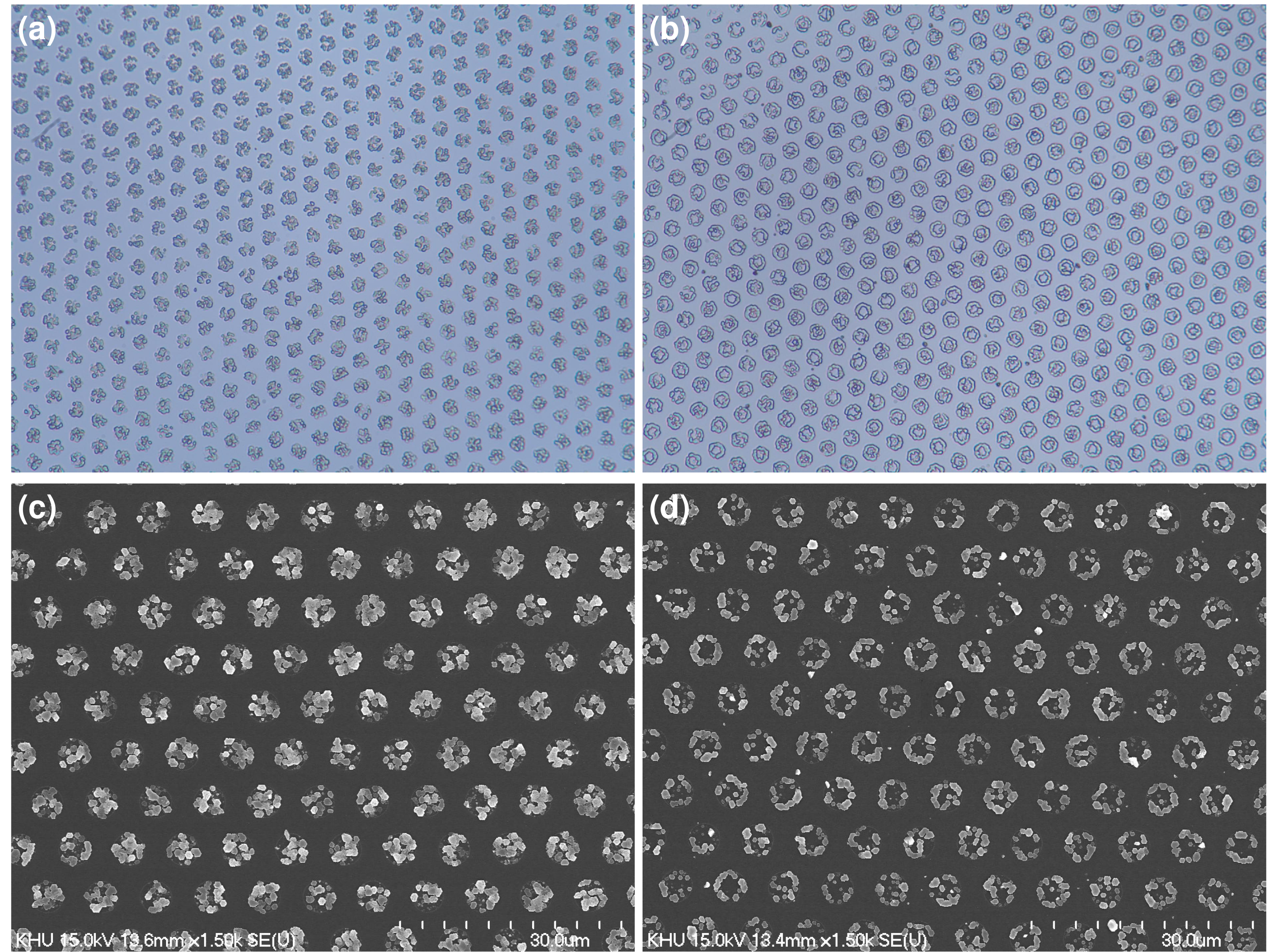}
\caption{Comparison of GaN nucleation after 21 s growth on sapphire substrates prepared at 700$^\circ$C under otherwise identical thermal conditions. (a,c) H$_2$-annealed sapphire without ammonia borane exposure and (b,d) sapphire exposed to ammonia borane for 15 min, resulting in a directly grown boron-compound surface layer or surface modification. Panels (a,b) show lower-magnification images, and panels (c,d) show corresponding higher-magnification images of the circular openings.}
\label{Annealing-vs-BN-film-effect}
\end{figure}

Figure~\ref{Annealing-vs-BN-film-effect} compares the initial GaN nucleation behavior on H$_2$-annealed sapphire at 700$^\circ$C and on sapphire exposed to ammonia borane under otherwise identical preparation conditions. The boron-compound-treated substrate exhibits a lower density of GaN nuclei and a smaller areal fraction of GaN than the H$_2$-annealed sapphire. Since the principal difference between the two samples is whether ammonia borane was introduced during the preparation step, this result indicates that the modified GaN nucleation behavior cannot be explained by thermal pretreatment alone. Instead, it supports the interpretation that ammonia borane exposure induces a precursor-derived surface layer or surface modification that suppresses the initial nucleation of GaN.

\subsection{Growth-temperature-dependent domain distribution}

\begin{figure} 
\includegraphics[width=1.0\columnwidth]{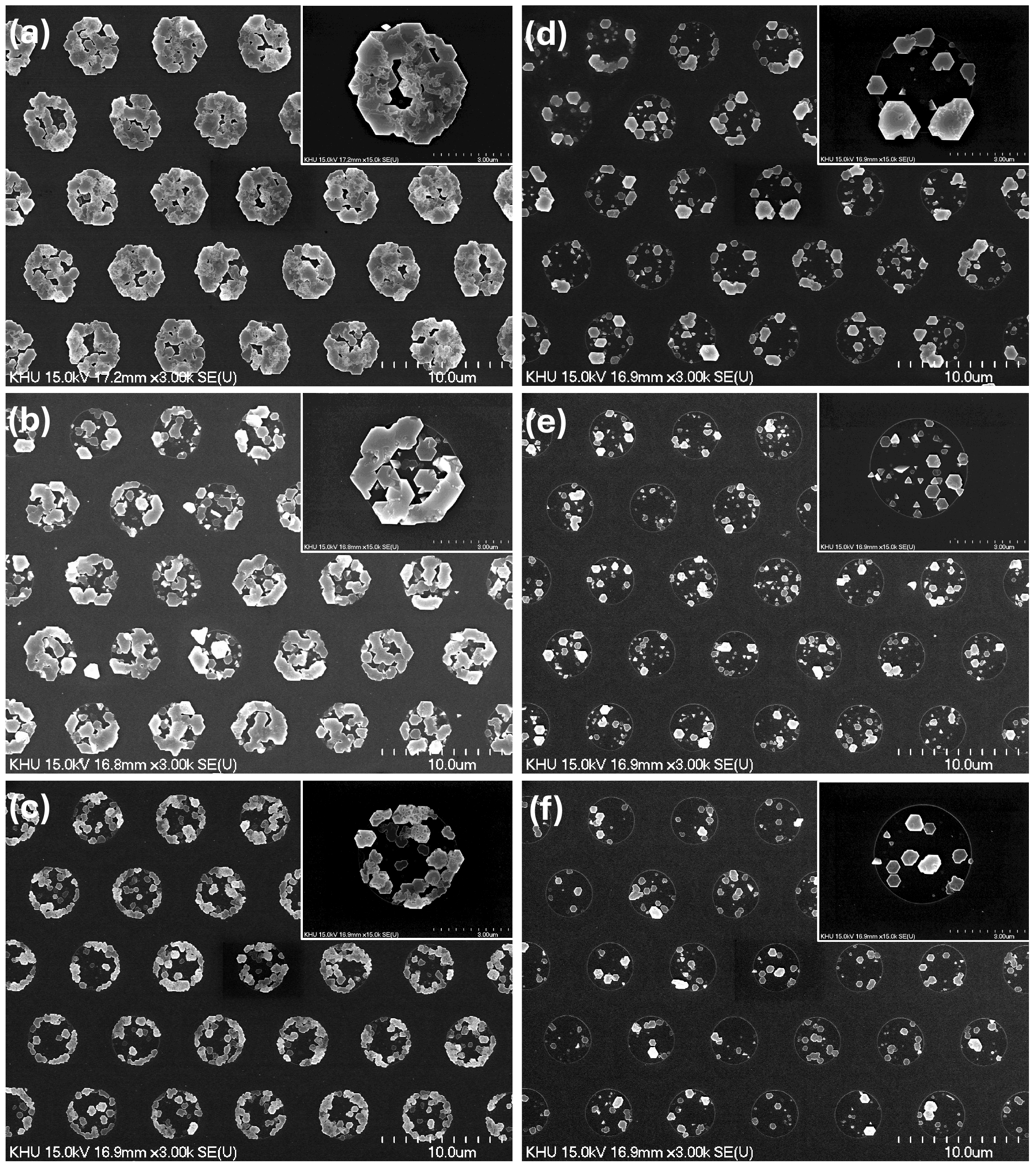}
\caption{SEM images of GaN domains grown for 21 s on \(c\)-plane sapphire substrates coated with directly grown boron-compound masks prepared at (a) 550$^\circ$C, (b) 650$^\circ$C, (c) 700$^\circ$C, (d) 750$^\circ$C, (e) 800$^\circ$C, and (f) 850$^\circ$C. The images show a growth-temperature-dependent change in both the GaN coverage and the preferred domain position within the circular SiO$_2$ openings.}
\label{temperature-dependent-BN-growth}
\end{figure}

Figure~\ref{temperature-dependent-BN-growth} shows the SEM images of GaN domains grown within circular SiO$_2$ openings after the formation of boron-compound masks at different temperatures. The GaN growth time was kept short to capture the early-stage nucleation and domain-growth behavior before complete coalescence. The SEM images reveal two clear temperature-dependent trends. First, as the boron-compound mask growth temperature increases, the overall GaN coverage within the openings decreases, and the domains become smaller and more spatially separated. Second, the preferred domain position also changes with mask growth temperature. Pronounced edge-biased domain distributions are observed for intermediate mask growth temperatures, particularly for the 700 and 750$^\circ$C samples, whereas higher mask growth temperatures lead to more inward and spatially sparse domain distributions. This temperature-dependent redistribution indicates that the growth temperature of the directly formed boron-compound mask strongly affects the intra-opening GaN nucleation landscape on patterned sapphire.

A plausible interpretation of this positional redistribution is that the effective accessibility of the underlying sapphire surface varies with the growth temperature of the boron-compound mask. Such accessibility may arise from local structural nonuniformities in the boron-compound layer that provide precursor-accessible pathways to the sapphire surface. For intermediate mask growth temperatures, the substrate-accessible regions appear to be more effective near the SiO$_2$ boundary, leading to pronounced edge-biased GaN nucleation within the circular openings. As the mask growth temperature increases, the effective substrate-accessible sites appear to shift or extend toward the interior of the openings, resulting in a more inward distribution of GaN domains. Thus, the observed positional change is best understood as a temperature-dependent redistribution of local sapphire accessibility within the patterned openings, rather than simply as a change in the overall nucleation probability.

\subsection{Analysis of domain distribution}

\begin{figure}
\includegraphics[width=1.0\columnwidth]{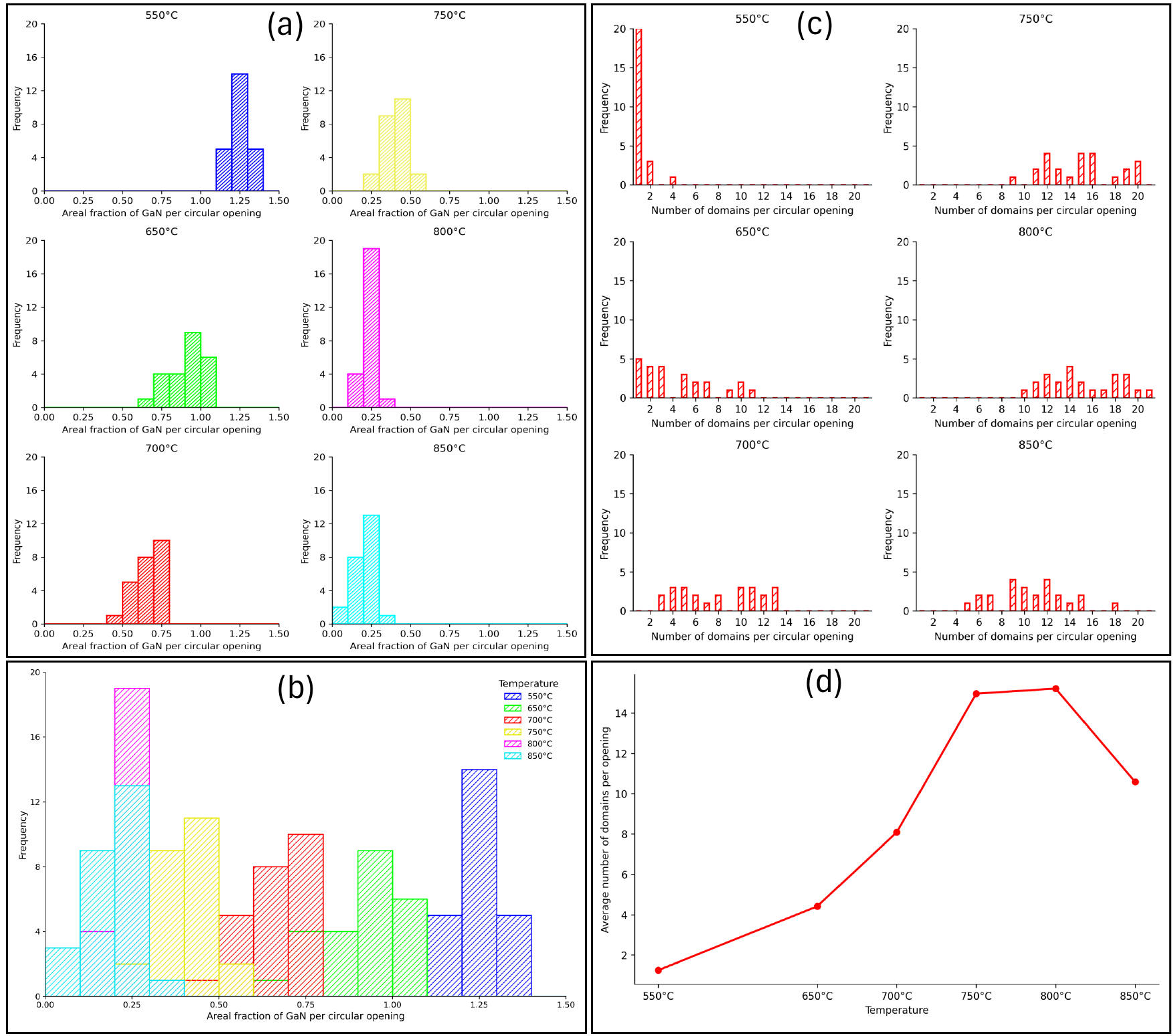}
\caption{Quantitative analysis of per-opening GaN domain statistics as a function of the boron-compound mask growth temperature. Statistics were obtained from 24 circular openings for each mask growth temperature. (a) Growth-temperature-dependent GaN areal fraction within circular SiO$_2$ openings. (b) Overlapped histograms of the GaN areal fraction for different mask growth temperatures. (c) Growth-temperature-dependent number distribution of GaN domains per circular opening. (d) Average number of visibly isolated GaN domains per circular opening as a function of mask growth temperature.}
\label{per-opening-scalar-quantities}
\end{figure}

To complement the qualitative SEM observations, we quantified two per-opening scalar metrics: the GaN areal fraction and the number of visible GaN domains within each circular opening. As shown in Fig.~\ref{per-opening-scalar-quantities}(a,b), the areal-fraction distributions shift systematically toward lower values as the growth temperature of the directly formed boron-compound mask increases. Openings prepared at 550--700$^\circ$C exhibit relatively high GaN coverage, whereas those prepared at 750--850$^\circ$C show substantially reduced areal fractions, with the 850$^\circ$C sample concentrated at the lowest values. At the same time, the domain-count statistics in Fig.~\ref{per-opening-scalar-quantities}(c,d) show a nonmonotonic dependence on the boron-compound mask growth temperature. The number of visibly separated GaN domains increases as the mask growth temperature is raised from 550 to 750--800$^\circ$C, but then tends to saturate and decreases at 850$^\circ$C. Thus, the experimental observation is not simply a monotonic decrease in nucleation activity. Rather, increasing the mask growth temperature first leads to lower GaN coverage while leaving a larger number of smaller and more spatially separated domains within the openings. At the highest mask growth temperature, however, the effective substrate-accessible regions appear to become sufficiently sparse or ineffective that both the GaN areal fraction and the number of visible domains are reduced.

This nonmonotonic change in the number of visible isolated domains should not be interpreted as a direct measure of the number of active nucleation sites. Instead, it reflects a competition between two effects: reduced lateral merging, which tends to increase the number of visibly separated domains, and reduced effective substrate accessibility, which tends to decrease the number of nucleation events. At intermediate-to-high mask growth temperatures, the effective substrate-accessible sites are inferred to become smaller or more widely separated, producing smaller GaN domains with reduced lateral impingement. As a result, more isolated domains can remain visible in SEM even though the total GaN areal fraction is reduced. In contrast, under the high-coverage conditions observed for the 550--700$^\circ$C mask-growth cases, closely spaced substrate-accessible sites or more extended accessible regions would promote rapid lateral growth and merging of neighboring domains, so that fewer isolated domains are counted after growth despite the larger GaN-covered area. At 850$^\circ$C, the further reduction in effective substrate accessibility appears to dominate over the reduced-merging effect, leading to a decrease in the number of visible domains.

\begin{figure}   
\includegraphics[width=1.0\columnwidth]{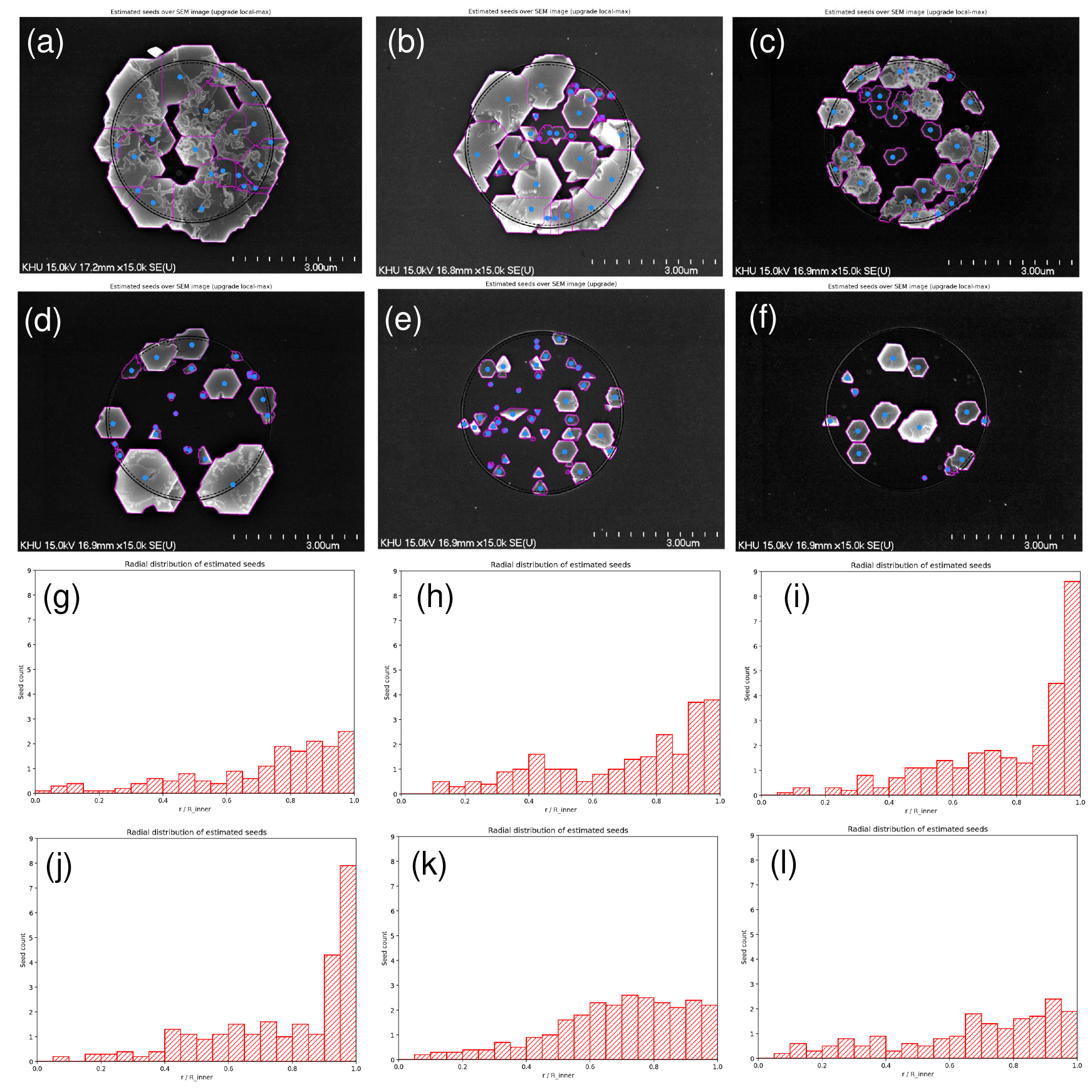}
\caption{Representative SEM images of GaN domains formed within circular SiO$_2$ openings after growth on sapphire coated with directly formed boron-compound masks prepared at different temperatures, together with the corresponding radial distributions of isolated domain centers.  Panels (a)--(f) correspond to mask growth temperatures of 550, 650, 700, 750, 800, and 850$^\circ$C, respectively, and panels (g)--(l) show the corresponding radial distributions. The upper panels (a)--(f) show SEM images with estimated centers of isolated GaN domains overlaid, and the circular boundary of each opening is indicated. The lower panels (g)--(l) show histograms (averaged over 10 circular openings) of the normalized radial coordinate \(r/R_{\mathrm{inner}}\), where \(r=0\) corresponds to the opening center and \(r/R_{\mathrm{inner}}=1\) corresponds to the inner boundary near the SiO$_2$ opening edge.}
\label{intra-opening-radial-distribution}
\end{figure}

To further examine how the directly formed boron-compound mask modifies the intra-opening nucleation landscape, we analyzed the radial positions of the centers of isolated GaN domains within each circular opening using the normalized coordinate \(r/R_{\mathrm{inner}}\), where \(r=0\) corresponds to the center of the opening and \(r/R_{\mathrm{inner}}=1\) corresponds to the inner boundary near the SiO$_2$ opening edge. As shown in Fig.~\ref{intra-opening-radial-distribution}, the domain-center distributions depend strongly on the boron-compound mask growth temperature. Pronounced edge-biased distributions, characterized by strong concentration near the high-\(r/R_{\mathrm{inner}}\) region, are observed for the intermediate mask growth temperatures, particularly 700 and 750$^\circ$C. At higher mask growth temperatures, the distributions broaden and extend toward smaller \(r/R_{\mathrm{inner}}\) values, indicating that the preferred positions of isolated domain centers shift inward and become more spatially sparse. These results show that the mask growth temperature affects not only the per-opening GaN coverage and domain statistics, but also the preferred radial location of GaN domains within each circular opening.

The radial distributions of isolated domain centers can be interpreted, to first order, as position-resolved proxies for the spatial distribution of effective substrate-accessible nucleation sites within the boron-compound mask. Because the analysis is based on the final SEM morphology after finite lateral growth, the extracted domain-center positions should not be regarded as direct observations of the initial atomic-scale nucleation sites. Nevertheless, for isolated or weakly merged domains, the domain-center position provides useful information on where GaN nucleation was most likely initiated within the opening. In this interpretation, the pronounced edge-biased distributions observed for the 700 and 750$^\circ$C mask-growth conditions suggest that substrate accessibility is locally enhanced near the SiO$_2$ boundary under these conditions. This enhanced accessibility may arise from local variations in the continuity, thickness, or defect structure of the boron-compound layer. As the mask growth temperature is further increased, the inward broadening of the radial distributions suggests that the effective substrate-accessible regions shift or extend toward the interior of the openings. Combined with the scalar-metric analysis, this positional analysis supports the conclusion that the boron-compound mask growth temperature reshapes the GaN nucleation landscape by modifying both the amount and the radial distribution of local sapphire accessibility.

Although direct visualization of the nanoscale substrate-accessible regions in the boron-compound masks would provide the most direct evidence, their density and spatial distribution could not be unambiguously resolved from the available AFM or TEM observations. Therefore, we do not assign the observed GaN nucleation behavior to a specific microscopic structure of the mask. Instead, the boron-compound layer is described in terms of effective substrate-accessible regions, which may include nanoscale openings, percolative through-pathways, locally thinner regions, or discontinuities. In this framework, the SEM-based GaN domain statistics and radial-position analysis provide a functional readout of local sapphire accessibility, while the kMC-type simulations, as shown in the next subsection, test whether changes in the density and radial distribution of such effective accessible regions can reproduce the experimentally observed trends.

\subsection{Kinetic Monte Carlo-type simulation}

\begin{figure}
\includegraphics[width=1.0\columnwidth]{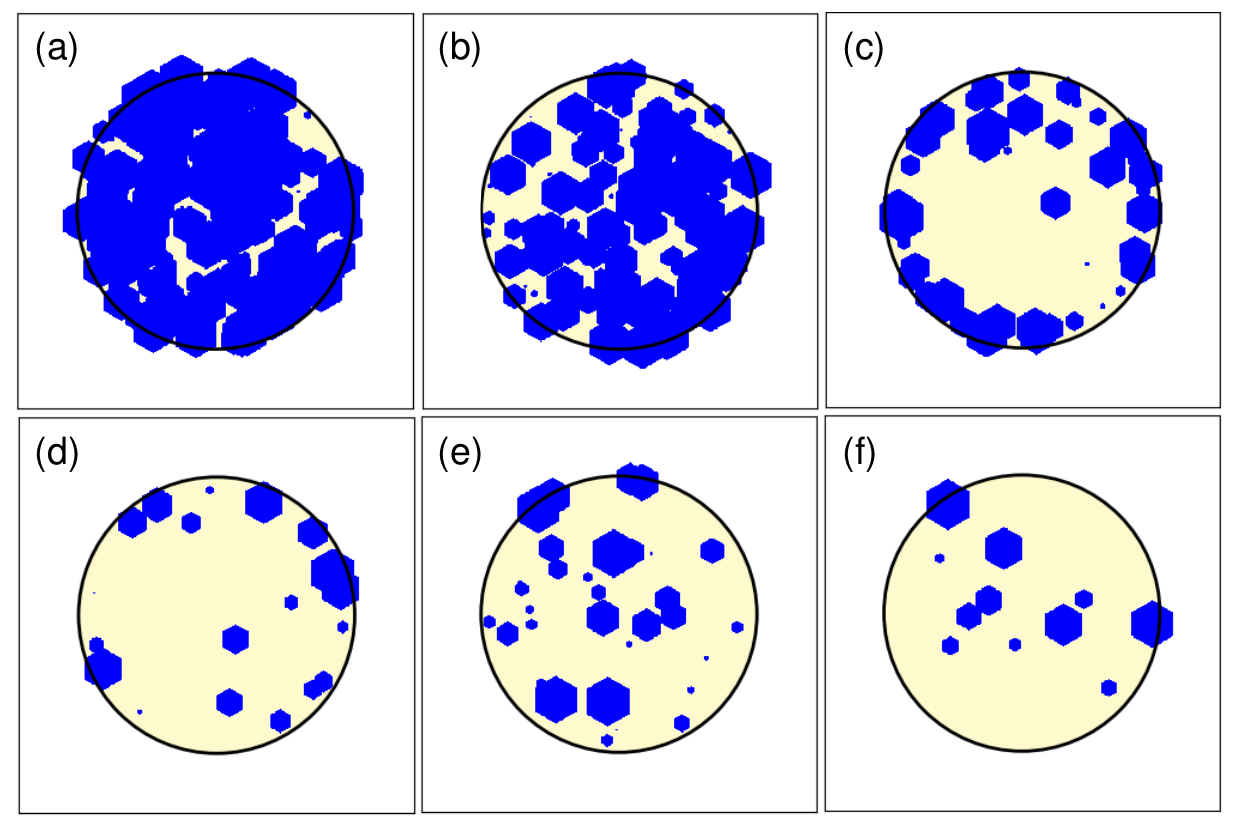}
\caption{
Kinetic Monte Carlo-type simulations of GaN nucleation and early-stage domain growth within a circular opening for different effective substrate-accessibility conditions. GaN nucleation was allowed only at effective substrate-accessible sites in the boron-compound mask, representing local regions where precursors can reach the underlying sapphire surface. The simulated morphologies are shown for representative parameter sets labeled according to the experimental mask-growth-temperature cases that they are intended to qualitatively emulate: (a) 550$^\circ$C, (b) 650$^\circ$C, (c) 700$^\circ$C, (d) 750$^\circ$C, (e) 800$^\circ$C, and (f) 850$^\circ$C. Blue regions denote GaN-covered areas within the opening. The effective accessible-site fractions, \(f_{\mathrm{acc}}\), were set to 0.77, 0.45, 0.165, 0.085, 0.080, and 0.025 for the 550, 650, 700, 750, 800, and 850$^\circ$C cases, respectively. For the 700 and 750$^\circ$C cases, a ring-biased distribution of accessible sites near the SiO$_2$ boundary was introduced with a target ring-to-inner site ratio of 9:1, whereas the other cases used spatially uniform accessible-site distributions. The nucleation rate per accessible site and the anisotropic lateral-growth kernel were kept fixed for all cases. The parameter \(f_{\mathrm{acc}}\) represents an effective accessible-site fraction used in the simulation and should not be interpreted as a directly measured microscopic opening fraction. The simulated morphologies are representative parameter sets intended to qualitatively reproduce the experimental trends.
}
\label{kMC}
\end{figure}

To further examine whether the experimentally observed temperature-dependent GaN nucleation behavior can be explained by changes in local sapphire accessibility, kinetic Monte Carlo-type simulations were carried out, as shown in Fig.~\ref{kMC}. In the model, GaN nucleation was allowed only at effective substrate-accessible sites within the circular opening. These sites represent local regions where GaN precursors can access the underlying sapphire surface through the boron-compound mask, without specifying the microscopic origin of that accessibility.

The representative mask-growth-temperature cases were simulated by varying two coarse-grained parameters: the effective accessible-site fraction, \(f_{\mathrm{acc}}\), and the radial distribution of the accessible sites. The values of \(f_{\mathrm{acc}}\) were set to 0.77, 0.45, 0.165, 0.085, 0.080, and 0.025 for the 550, 650, 700, 750, 800, and 850$^\circ$C cases, respectively. These values should not be interpreted as directly measured microscopic opening fractions, but as effective parameters describing the probability that local regions of the boron-compound mask permit precursor access to the sapphire surface. The nucleation rate per accessible site and the anisotropic lateral-growth kernel were kept fixed for all cases.  These representative parameter sets were not intended to provide a unique one-to-one reconstruction of the microscopic mask morphology at each temperature. Rather, they were used to separate two effects: the total effective accessible-site fraction and the radial bias of accessible sites.

For the 550 and 650$^\circ$C cases, the relatively high values of \(f_{\mathrm{acc}}\) produce a large number of nuclei and extensive lateral coalescence, resulting in high GaN-covered area fractions within the opening. For the 800 and 850$^\circ$C cases, the lower values of \(f_{\mathrm{acc}}\) reduce the number of accessible nucleation positions, leading to smaller and more spatially separated GaN domains. This reproduces the experimentally observed transition from high-coverage, laterally merged domains to sparse isolated domains.

In addition to changing the total effective accessible-site fraction, the radial distribution of accessible sites was varied for the 700 and 750$^\circ$C cases. For these cases, the accessible sites were distributed preferentially near the SiO$_2$ opening boundary using a ring-biased distribution with a target ring-to-inner site ratio of 9:1. This radial bias produces preferential nucleation near the opening edge, demonstrating that the experimentally observed edge-biased nucleation can be reproduced by changing the spatial distribution of effective substrate-accessible sites, rather than by changing only their total fraction.  The uniform distributions used for the 550 and 650$^\circ$C cases were not intended to exclude edge-biased accessibility in the actual samples. Rather, because the high \(f_{\mathrm{acc}}\) values in these cases already led to extensive lateral merging, the radial bias of individual accessible sites could not be uniquely extracted from the final morphology. Ring-biased distributions were therefore introduced only for the intermediate cases to demonstrate the effect of radial accessibility bias under reduced-coalescence conditions.

The simulations therefore reproduce the essential experimental trends: high GaN coverage when the effective substrate-accessible fraction is large, sparse isolated domains when the accessible-site fraction is reduced, and edge-biased nucleation when the accessible sites are radially concentrated near the SiO$_2$ boundary. These results support the interpretation that the mask growth temperature reshapes the intra-opening GaN nucleation landscape by modifying both the amount and the radial distribution of local substrate-accessible regions, rather than simply changing a uniform nucleation probability over the entire opening.

\subsection{Chemical bonding states of boron-compound masks}

\begin{figure}
\includegraphics[width=1.0\columnwidth]{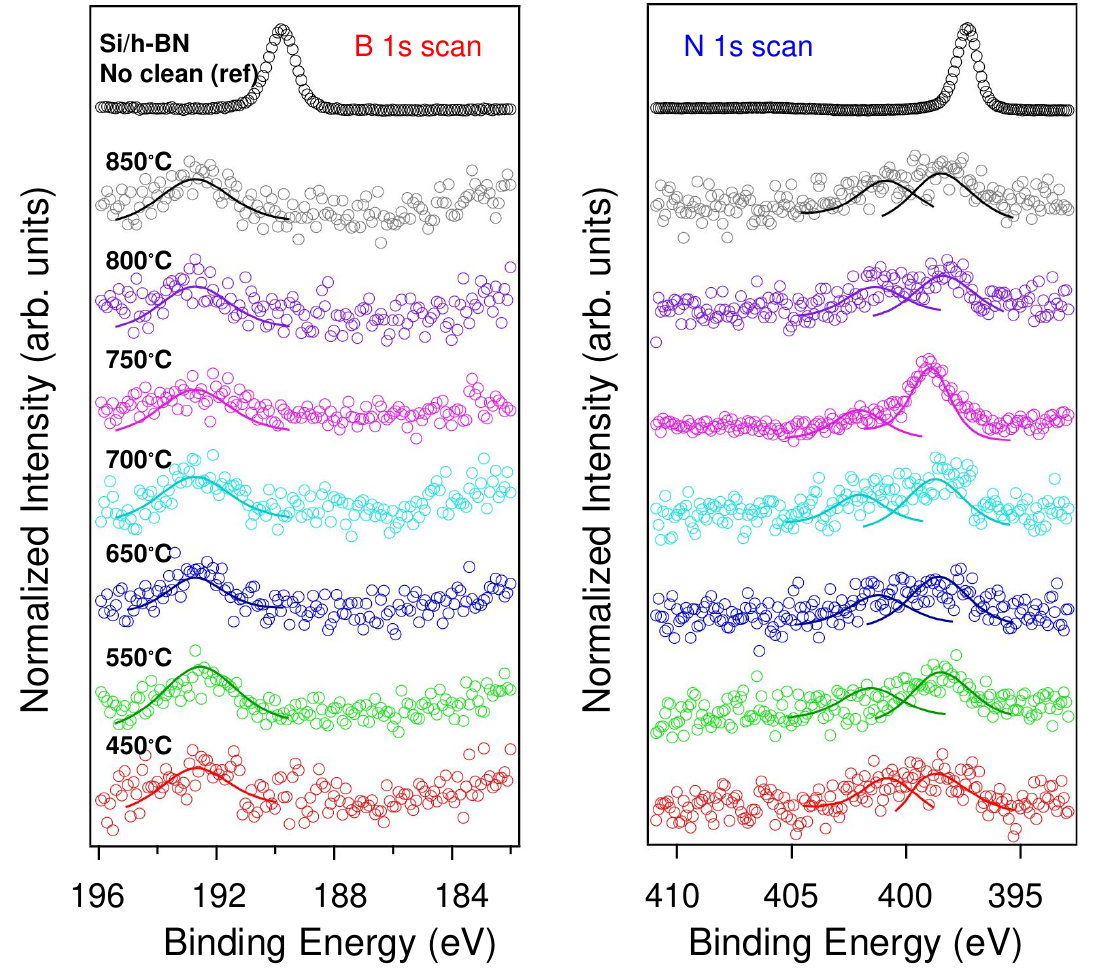}
\caption{XPS B 1s and N 1s core-level spectra of boron-compound masks grown at different temperatures. Spectra from films grown at 450, 550, 650, 700, 750, 800, and 850$^\circ$C are shown together with a Si/h-BN reference spectrum for comparison. The directly grown boron-compound masks exhibit broader B 1s and N 1s features than the reference, indicating mixed boron-containing bonding states rather than a single phase-pure hBN bonding environment.}
\label{XPS-spectra-of-BN-grown-at-different-temperatures}
\end{figure}

Because the boron-compound masks in this work were grown at temperatures substantially lower than those typically used for high-quality hBN synthesis, their chemical state cannot be assumed to correspond to phase-pure hBN. X-ray photoelectron spectroscopy (XPS) was therefore performed to examine the bonding states of the directly grown boron-compound masks. Figure~\ref{XPS-spectra-of-BN-grown-at-different-temperatures} shows the B 1s and N 1s core-level spectra of films grown at different temperatures, together with a Si/h-BN reference spectrum for comparison.  The XPS measurements were performed on additional boron-compound mask samples prepared under similar, but not identical, conditions to those used for the SEM-based nucleation analysis. Therefore, the XPS data are used here to identify the general chemical bonding character of the mask material, rather than to establish a temperature-by-temperature correlation with the GaN domain distribution.

The Si/h-BN reference exhibits relatively sharp B 1s and N 1s features at binding energies characteristic of B--N bonding. In contrast, the spectra of the directly grown boron-compound masks are substantially broader, indicating that the films do not consist of a single, well-defined hBN bonding environment. The B 1s signals of the mask samples are distributed mainly around 191--193 eV, with a pronounced contribution on the higher-binding-energy side compared with the Si/h-BN reference. The lower-binding-energy side can be associated with BN-related bonding, whereas the higher-binding-energy contribution may include defective or \(sp^3\)-rich BN-related bonding states as well as oxidized boron species.\cite{Koepke-CM-28-4169,Ong-JAP-95-3527} The N 1s spectra also exhibit broad features around 398--401 eV, suggesting the presence of nitrogen-containing bonding states. However, the broadness and sample-to-sample spectral variation do not allow unambiguous assignment to phase-pure hBN.\cite{Koepke-CM-28-4169}

Therefore, the XPS results support the use of the term ``boron-compound mask'' throughout this work. The films are best described as precursor-derived boron-containing layers with mixed bonding states, including oxide-related and nitride-related components, rather than as high-quality or phase-pure hBN. This chemical complexity is consistent with the proposed role of the mask as a spatially nonuniform layer that modifies the effective accessibility of the underlying sapphire surface during GaN nucleation. In this context, the relevant function of the film is not only its crystallographic identity, but also its local continuity, thickness, defect structure, and substrate accessibility.

\subsection{Crystallographic alignment of GaN}

\begin{figure}
\includegraphics[width=0.75\columnwidth]{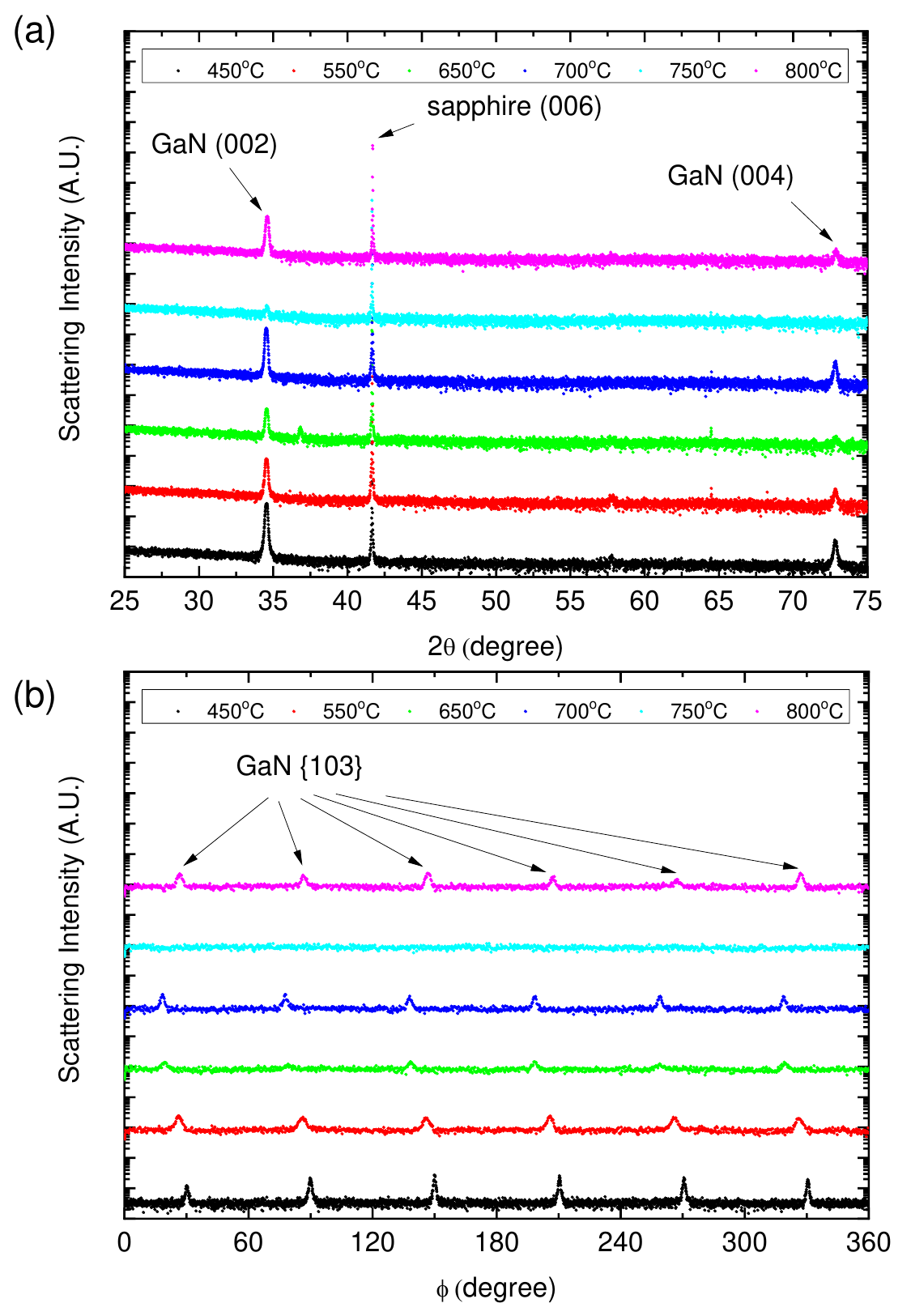}
\caption{XRD characterization of GaN grown on additional boron-compound-coated sapphire samples prepared under conditions similar to those used for the GaN nucleation study. (a) \(2\theta\)-\(\omega\) scans showing GaN (002) and (004) reflections together with the sapphire (006) reflection. (b) \(\phi\)-scan of the GaN (103) reflection showing in-plane alignment. These samples were not identical to those used for the SEM-based nucleation statistics and are used here as complementary evidence for the crystallographic alignment of GaN with the underlying sapphire substrate.}
\label{XRD}
\end{figure}

Figure~\ref{XRD} presents XRD measurements obtained from additional GaN/boron-compound/sapphire samples prepared under growth conditions similar to those used for the nucleation-position analysis. Although these samples are not identical to those used for the SEM-based statistical analysis, they provide complementary information on the crystallographic relationship between GaN and the underlying sapphire substrate. The \(\theta\)-\(2\theta\) scans exhibit GaN (002) and (004) reflections together with the sapphire (006) reflection, indicating that the GaN domains are predominantly \(c\)-axis oriented. In addition, the \(\phi\)-scan of the GaN (103) reflection shows well-defined in-plane alignment, suggesting that GaN grown in this material system can retain an epitaxial relationship with the sapphire substrate.  Because these XRD measurements were performed on additional samples prepared under similar, but not identical, conditions, the temperature labels in Fig.~\ref{XRD} should not be interpreted as providing a one-to-one correlation with the SEM-based domain statistics. The data are used only to confirm that GaN grown in this material system can retain crystallographic alignment with sapphire.

This crystallographic alignment is consistent with nucleation at local substrate-accessible regions in the boron-compound mask, such as discontinuities, locally thinner regions, or through-pathways, rather than with completely random nucleation on an intact and fully continuous mask surface. Although nucleation assisted by the boron-compound surface cannot be completely excluded, the observed out-of-plane and in-plane alignment of GaN with sapphire is difficult to explain by random nucleation on a chemically mixed and structurally disordered boron-compound layer alone. This supports the view that local access to the underlying sapphire substrate plays an important role in determining the GaN nucleation sites. However, because the XRD samples are not identical to those used for the SEM-based nucleation statistics, this result should be regarded as complementary support for the substrate-accessibility picture rather than direct proof of the microscopic origin of each GaN domain observed in Fig.~\ref{temperature-dependent-BN-growth}.

\section{Conclusion}

In this work, we investigated how the growth temperature of directly grown ammonia-borane-derived boron-compound masks affects GaN nucleation on SiO$_2$-patterned \(c\)-plane sapphire. The control experiment using H$_2$-annealed sapphire showed that the change in GaN nucleation behavior cannot be explained by thermal pretreatment alone, but is associated with ammonia-borane-induced formation of a precursor-derived boron-compound surface layer or surface modification. SEM-based analysis revealed that increasing the mask growth temperature reduces the GaN areal fraction, while the number of visibly isolated GaN domains changes nonmonotonically, increasing up to 750--800$^\circ$C and decreasing at 850$^\circ$C.  Radial-position analysis further showed that the preferred domain location varies with mask growth temperature, with pronounced edge-biased distributions observed for the 700 and 750$^\circ$C mask-growth conditions and more inward distributions observed at higher mask growth temperatures.

These results indicate that the growth temperature of the boron-compound mask reshapes the effective accessibility of the underlying sapphire surface within patterned openings. This change may originate from temperature-dependent variations in mask coverage, thickness, continuity, defect structure, or local precursor-access pathways, including possible through-pathways. Kinetic Monte Carlo-type simulations support this interpretation by showing that changes in the density and radial distribution of effective substrate-accessible sites can reproduce the observed changes in GaN coverage and nucleation position. Overall, this work demonstrates that the growth temperature of directly formed boron-compound masks provides a practical process parameter for controlling the intra-opening GaN nucleation landscape on patterned sapphire.

\section{acknowledgements}
This work was supported by the National Research Foundation of Korea (NRF) grant funded by the Korea government (MSIT) (RS-2021-NR060087, RS-2023-00240724) and through Korea Basic Science Institute (National research Facilities and Equipment Center) grant (2021R1A6C101A437) funded by the Ministry of Education.


%

\end{document}